\documentclass{article}

\usepackage{arxiv}

\usepackage[utf8]{inputenc} 
\usepackage[T1]{fontenc}    
\usepackage{hyperref}       
\usepackage{url}            
\usepackage{booktabs}       
\usepackage{nicefrac}       
\usepackage{microtype}      

\usepackage{amsmath}

\usepackage{lineno,hyperref}
\usepackage{siunitx}
\modulolinenumbers[5]

\usepackage[table]{xcolor}
\usepackage{array}
\usepackage{graphicx}
\usepackage[misc,geometry]{ifsym} 
\usepackage{lipsum}
\usepackage{subfig}
\usepackage{calc}
\usepackage{verbatim}
\usepackage{amssymb}
\usepackage{amstext}
\usepackage{amsthm}
\usepackage{mathtools}
\usepackage[ruled]{algorithm2e}
\usepackage{multirow}
\usepackage{mathrsfs}
\usepackage{mathtools}
\usepackage{tabularx}
\usepackage{amsfonts}
\newcommand{\cmt}[1]{ }
\newcommand{\suchthat}{\;\ifnum\currentgrouptype=16 \middle\fi|\;}
\newcommand{\?}{\stackrel{?}{=}}

\newlength{\Oldarrayrulewidth}
\newcommand{\Cline}[2]{%
  \noalign{\global\setlength{\Oldarrayrulewidth}{\arrayrulewidth}}%
  \noalign{\global\setlength{\arrayrulewidth}{#1}}\cline{#2}%
  \noalign{\global\setlength{\arrayrulewidth}{\Oldarrayrulewidth}}}

\title{Robust Node ID assignment for Mobile P2P Networks}

\author{
  Sumit Kumar Tetarave \\
  Department of Computer Science \& Engineering\\
  Indian Institute of Technology Patna\\
  Bihar, India 801103 \\
  \texttt{sktetarave@iitp.ac.in} \\
   \And
 Somanath Tripathy \\
  Department of Computer Science \& Engineering\\
  Indian Institute of Technology Patna\\
  Bihar, India 801103 \\
  \texttt{som@iitp.ac.in} \\
}

\begin{document}
\maketitle

\begin{abstract}
The advancement of portable mobile wireless devices such as smart-phones, PDA, etc., brought mobile peer-to-peer (P2P) as an extension of traditional P2P networks to provide efficient, low-cost communication among them in a cellular network. 
It is challenging to assign a unique identifier to each user, as an adversary can target to disrupt the P2P system, by carefully selecting user IDs or obtaining many pseudo-IDs. 
This work proposes a robust node-ID assignment mechanism for secure peer joining in mobile P2P system called PJ-Sec. PJ-Sec facilitates to generate nodeID for a joining peer by a collaborative effort of an existing peer (within the vicinity) and pre-selected vicinity head. 
PJ-Sec is formally analyzed using AVISPA model checker and found to be attack resistant. 
\end{abstract}


\section{Introduction}
\label{intro}
Annual Visual Networking Index (VNI) Forecasts,
smart-phones will surpass 86\% of mobile data traffic by 2021, and it will grow at a compound annual growth rate (CAGR) of 47\% from 2016 to 2021~\cite{cisco2017cisco}. To reduce network traffic in a cellular network, an alternative solution is to introduce peer to peer (P2P) overlay communication through device-to-device (D2D)/Bluetooth/WiFi. Many research works have been carried out to collaborate P2P overlay with mobile infrastructure (5G/LTE-A/3G, etc.) to provide scalable and efficient data sharing such as C-Chord~\cite{cchord2012}, V-Chord~\cite{tetarave2018v}. 

On the other hand, most of
the P2P overlays are developed without considering the security aspects during the new identifier (ID) assignment. 
Generally, the overlays generate peer ID locally using hash computation over IP addresses (like Chord~\cite{chord2003}, Kademlia~\cite{maymounkov2002kademlia}), or randomly by client software as in Pastry~\cite{Rowstron:row}, or by their assigned clusters which are selected by cluster members as in CAN~\cite{ratnasamy2001scalable}.

The underlying assumption in these overlays is that its members (smart-phones, PDAs, etc.) behave honestly, which would be hard in a wireless environment. It is observed that the existence of anonymous peers and lack of central authority make the overlay vulnerable. Best way to develop a secure overlay is to restrict the malicious peer during the bootstrap process. Thus, secure node joining is considered to be the topmost priority to develop the secure routing primitives such as secure maintenance of routing table and routing messages as discussed in ~\cite{wallach2003survey,lee2018rfid}.



The centralized secure nodeID assignment mechanisms including a Robust Identity Assignment Protocol for P2P overlays (RIAPPA) ~\cite{caubet2014riappa}, and Identity Assignment Protocol (IAP) ~\cite{caubet2013securing} suffer from a single point of failure. On the other hand, decentralized solutions like ~\cite{khan2015midep,dinger2006defending} uses collaborative efforts of some existing peers without requiring a centralized server. But, these security protocols need high computation and communication overhead, so not suitable to use in mobile or cellular systems. 

Meanwhile, a mobile or cellular DHT overlay can be framed to simplify the P2P applications like file-sharing. Moreover, the whole network can be divided into several clusters/vicinity with at least one serving super-peer/cluster head in each vicinity.



This work proposes a secure node joining protocol tailoring to super-peer based cellular overlays. 
This mechanism assigns nodeID to a new peer as a collaborative effort of an existing overlay member ( a friend peer) and its serving super-peer, mitigating Sybil, Eclipse, and  man-in-the-middle (MITM) attacks.. 

The rest of the paper is organized as follows. Related work is presented in Section \ref{relatedWork}. 
Section \ref{sysModel} illustrates the system model, points out its vulnerability, and design requirement for ID assignment protocols. In Section \ref{secModel}, we present the proposed secure peer assignment protocol (PJ-Sec). Design requirement analysis, security analysis, and formal verification are discussed in Section \ref{analysis}. Implementation and result analysis of PJ-Sec are performed in  Section~\ref{dis}. Finally, we present the concluding remarks in Section \ref{con}. 

\section{Related Work}
\label{relatedWork}
There are several secure P2P nodeID assignment techniques proposed in past, which can be categorized as centralized and distributed key-based solutions.  
\subsection{Centralized nodeID generation}
 In this approach, new nodes are generated and certified through a certificate authority (CA) as in ~\cite{Castro:2002:SRS:844128.844156,douceur2002sybil,Srivatsa_vulnerabil}.
Due to the centralized management of keys, it provides a unique identity for a new node. This feature makes it resistant against Sybil identities. These CAs are not responsible for distributing the nodeIDs uniformly in a virtual space, but the assigned IDs are properly bound with their assigned certificate through respective CA. Using these certificates, CAs can visualize illegal actions of nodes and revoke their certificates to eliminate the compromised nodeIDs. Certification revocation is costly due to processing and administrative overheads. 

For generating nodeID, ~\cite{butler2009leveraging,aiello2008tempering,aiello2011identity} introduced joint authentication process such as identity based encryption (IBE) through trusted authority. In this, the hash function is computed over the signed public key, or human interaction with CAs to improve the efficiency. It helps CAs in the nodeIDs authentication process.  

Caubet \textit{et al.} in 
~\cite{caubet2013securing} proposed an implicit certificate based solution. Implicit certificates ignore the issuer's signature and thus, makes it less computing with reduced size. 
In this protocol,  new nodeIDs are generated by  
collaboration with CA. To preserve user's anonymity with traceability feature,  RIAPPA~\cite{caubet2014riappa}, was proposed. RIAPPA uses two trusted third parties (TTP) and the IDs are authenticated by an external TTP using their real-world digital certificate, while the internal TTP manages their node IDs jointly.  

Vinayagam \textit{et al.} in ~\cite{vinayagam2018secure} presented a  restricted identity-based proxy re-encryption mechanism to mitigate forge IDs in P2P overlays. Each peer has to register its ID to a proxy server, after joining into the overlay. After registration, the proxy generates a signature for each node, which can be verified during communication at both the (sender and receiver) ends. Thus,  forge nodeIDs can be detected. 

\subsection{Distributed nodeID generation}

The existing mechanisms in this category simplify new node joining in a distributed overlay. Dinger \textit{et al.}  ~\cite{dinger2006defending} verified the generated nodeID by a certain number of existing nodes (\{E\}) in the overlay. 
For unauthorized ID, the certificate is revoked by $N_j (\in \{E\})$ to protect from Sybil IDs. Certificate revocations would be unsuitable for a mobile phone due to energy consumption and tariff for the Internet.

Khan \textit{et al.} ~\cite{khan2015midep} proposed a secure identity establishment protocol, called MIDEP, for P2P distributed peers. It establishes a secure non-traceable public identity with the help of peer collaboration to avoid identity shadowing and MITM attacks. In this, a unique user ID is divided into different parts and some random parts are selected to send to the collaborators. Each collaborator receives a unique ID part for further processing. A temporary public ID is constructed after receiving a threshold number of processed (response from the collaborators) ID parts.     

Castro \textit{et al.} ~\cite{Castro:2002:SRS:844128.844156} presented a cryptographic puzzle for binding generated ID with its corresponding node IP address. In this, each user chooses a key pair to make first p bits to zero after hashing the public key.  This computational challenge limits the nodeIDs to generate Sybil IDs. Further, Rowaihy \textit{et al.} ~\cite{rowaihy2007limiting} proposed a series of cryptographic puzzles to limit the Sybil attack. They used refreshment mechanism for assigned challenges to avoid adversary recalculation of the given challenge. To reduce computational cost during refreshment, Costa \textit{et al.} ~\cite{da2012identity} introduced an adaptive mechanism for generating computational puzzles which reduce forge ID attacks but not completely. Fang \textit{et al.}~\cite{fang2017self} developed a multi-criteria
fuzzy decision-making model  considering the dynamic of networks. It is self-adaptive and uses game theory concept to predict trust in the fuzzy and complex environment. In this, a newly joined node is assigned (by the existing nodes) a nonessential data during the test-period to build the trust. It helps to isolate the node if the trust values are not reached to a threshold during a specific period.

  Avramidis \textit{et al.} ~\cite{avramidis2012chord}, proposed distributed trust infrastructure for Chord-based P2P overlay to reduce computational cost during new nodeID generation and verification. In this, nodes maintain a set of certificates as a tag of each revocation without executing the actual revocation process. To improve the distributed trust infrastructure, Shi \textit{et al.} ~\cite{shi2013sybilshield} proposed a human-established trust model, which identifies Sybil IDs with the help of human feedback. These models take several communications to build trust among nodes including some possibility of forge ID generation. Inspired by the human-established trust model,
Xianfu \textit{et al.}~\cite{meng2018getrust} developed a guarantee-based trust model for Chord-based P2P networks in which, a joining peer evaluates all the eligible service peers using reputation. The new node joins with the service provider having a higher reputation.   However, the presence of a malicious service provider would influence the trust establishment for local node-ID generation process~\cite{brown2018security}, leading to a cascade effect. 

\section{System model, Adversary model, and Design Requirements}
\label{sysModel}
In this section, we discuss  the operational structure of the super-peer based P2P overlay architecture and state the goal of an adversary.

\subsection{System Model}
\begin{figure*}[!ht]
   \centering
     \subfloat[An Overlay with underlay connection in mobile P2P. \label{fig:threeCom}]{%
       \includegraphics[width=0.7\textwidth]{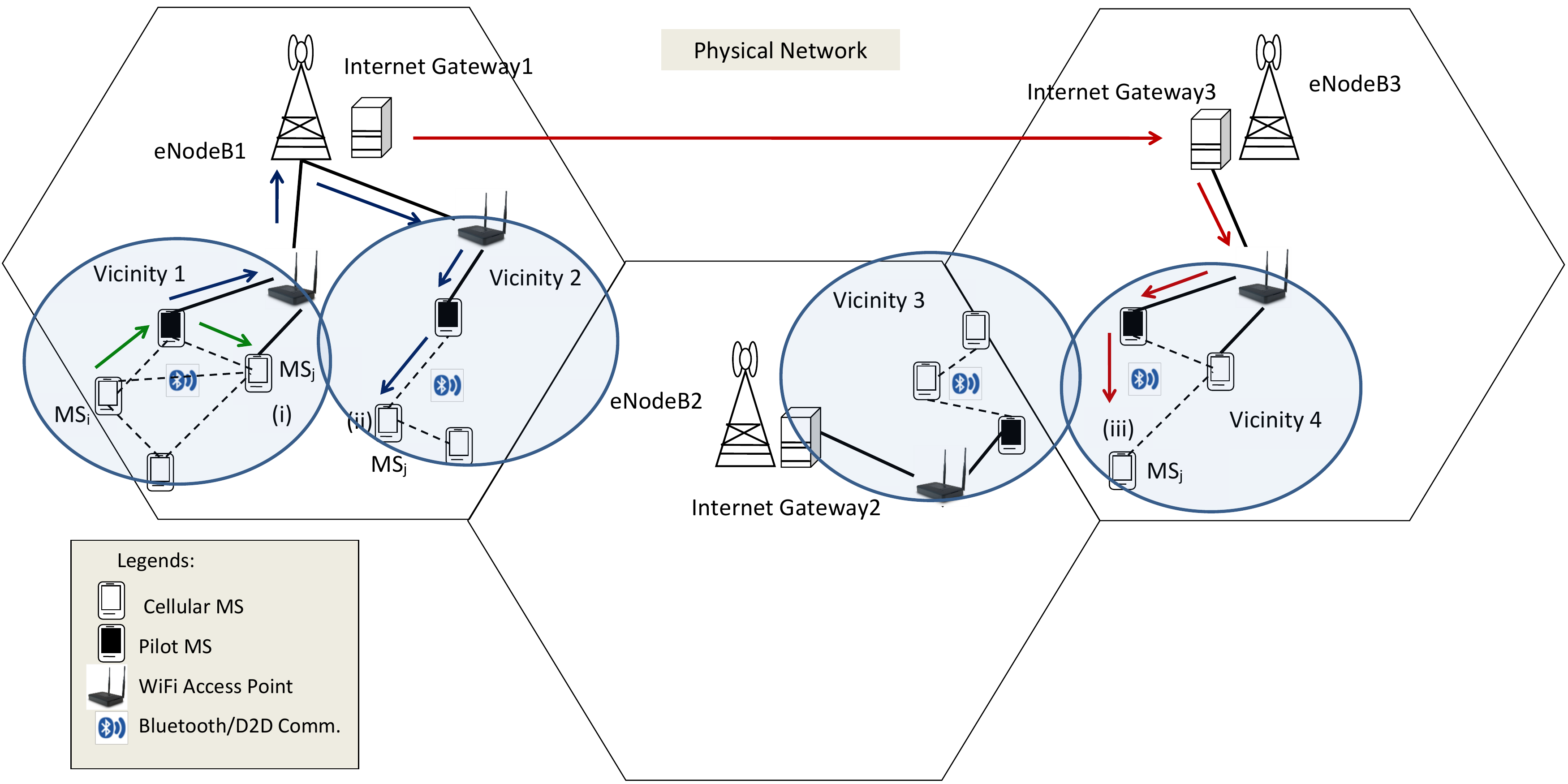}
     } \\
     \subfloat[An Overlay scenario with b = 2-bit, p = 3-bit and h = 5-bit \label{fig:LUUnderlay}]{%
       \includegraphics[width=0.5\textwidth]{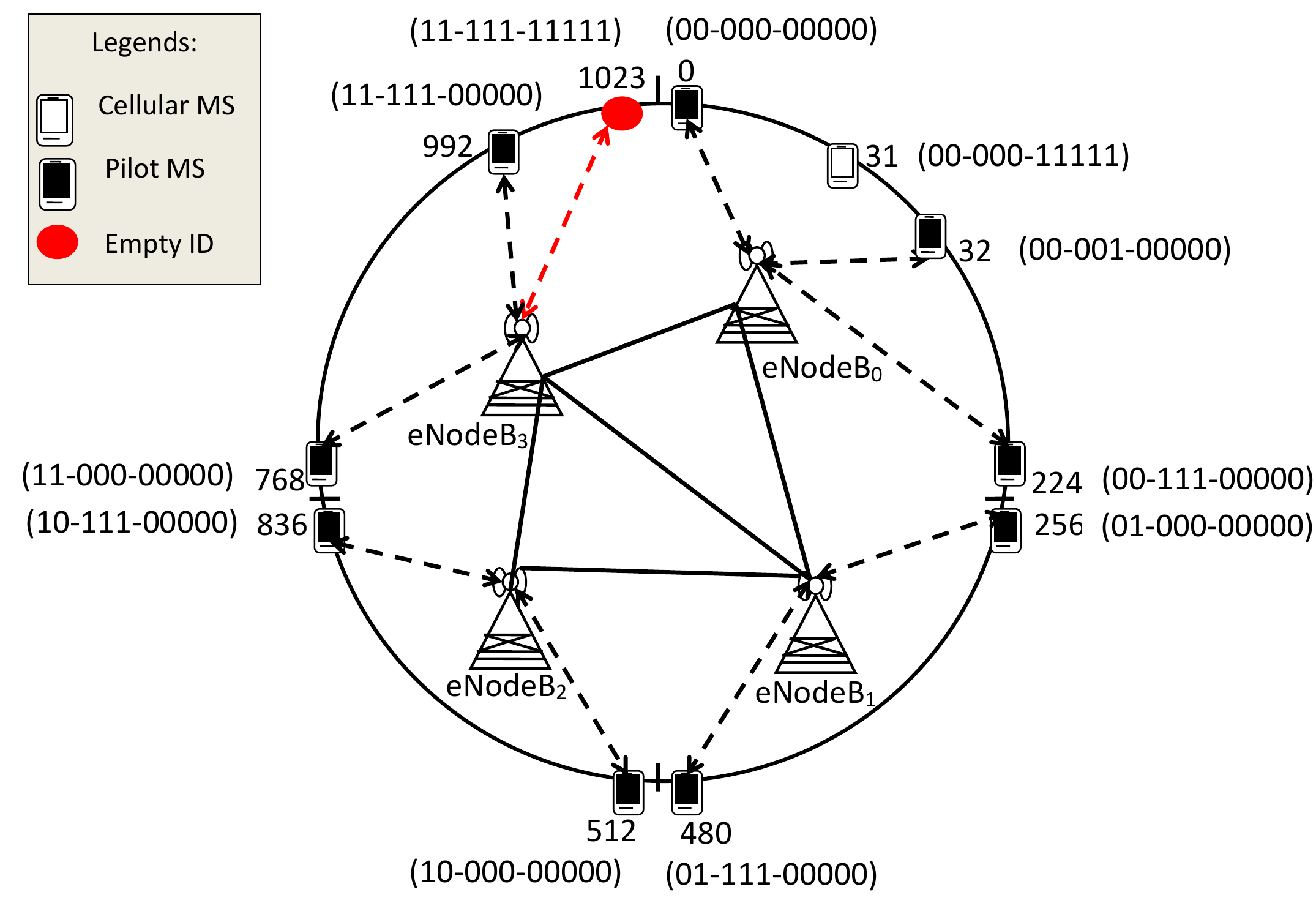}
     }
     \caption{Super-peer based mobile P2P overlay.}
     \label{fig:overlayConst}
   \end{figure*}

Consider a system like 
(the current) Gnutella~\cite{gnutella} network, which consists of super-peers~\cite{yang2003designing}. In cellular networks, these super-peers consume Internet connection and wish to serve other peers (smart-phones) through WiFi/Bluetooth/D2D connectivity as in ~\cite{tetarave2018v,cchord2012}. The overlay members (peers and super-peers) are connected within their communication range forming a cluster/vicinity as depicted in Figure ~\ref{fig:threeCom}. Vicinity 1 and 2 are managed by separate super-peers under the same  eNodeB 1 region, while vicinity 3 and 4 are in eNodeB 2 and eNodeB 3 respectively. A mobile station ($MS_i$) communicates with the target node ($MS_j$), which is within the range of WiFi connected super-peers (intra-region), within the  same eNodeB (intra- region) or another eNodeB region (inter-region).

Each member has a unique overlay ID and constructs an overlay structure as shown in Figure~\ref{fig:LUUnderlay}. In this, each ID is assigned with an $m$-bit DHT overlay ID. The $b$ most significant bit of $m$-bit specifies the corresponding $eNodeB$. Next, $p$-bit specifies the associated super-peer and remaining $h$-bit specifies the node under the super-peer. The ID assignment process of a new node is performed by a super-peer. The joining request may reach at super-peer through the existing peer(s).

%

%

\subsection{Adversary model}
We assume that vicinity head (or super-peer) are chosen periodically and on demand, by the majority (voting).  The eNodeB and super-peer are assumed to be trusted. 
The adversary is considered to be a probabilistic polynomial time (PPT) which cannot break (existing standard) cryptographic mechanisms. The primary goal of the adversary is to generate multiple fake IDs called Sybil IDs targeting to perform different attacks and control the network.

\begin{itemize}
\item Sybil Attack:
Maintaining multiple forge IDs to perform malicious activities is referred to as Sybil attack~\cite{douceur2002sybil}. These artificial nodeIDs behave like a genuine nodeID of a network and try to spread themselves to take control over the network.
\item Eclipse Attack: Purpose of this attack is to make a group of nodes isolated from rest of the network. In this, a genuine node is forced to communicate with others through adversary nodes.
\item MITM Attack: Here, an intermediate node tries to manipulate or handle secretly two-party communication without their concern. 
\end{itemize}


\subsection{Design Requirements}
A secure and robust identity assignment mechanism is desired to have the following features. 
\begin{enumerate}
\item Uniqueness: This requirement is specific to reduce the Sybil attack to provide unique nodeID to each real identity. The solution should restrict adversaries to generate multiple IDs.

\item Stability: To maintain its own reputation, this requirement enforces that a node should not change its own nodeID itself. The collaboration effort while node IDs generation would perfect in stability. 
  
\item Joint Management:  This requirement is necessary to avoid Eclipse attack. In this, nodeIDs must be issued with collaboration  among few peers. 

\item Verifiable: In which, member nodes can be able to check whether the nodeID and generated certificate are appropriately bounded or not. 

\item Traceability: This requirement facilitates that a misbehavior node should be traced with its given identity and generated nodeID.  
If a user does some malicious actions within its overlay, then the proposed scheme should be able to trace back the corresponding user at a later point of time and could be able to match the nodeID with the user’s unique identity.

\item Revocability: This requirement states that if any malicious activity is detected, the certificate of the corresponding user can be revoked. 

\item Uniformity: This requirement is necessary to distribute a load of an overlay to generate uniformly nodeIDs. NodeID should be assigned from a uniform distribution of overlay IDs within an allocated ID space. 

\item Scalability: It provides a large number of nodes to associate in the overlay. 

\item Efficient: This requirement is necessary to save communication cost, computational cost, and energy consumption of overlay mobile members. 

\end{enumerate}
\section{PJ-Sec: The Proposed Scheme}
\label{secModel}

The basic idea of  PJ-Sec is that a new node has to know a friend (bootstrap) node which would forward the joining request to the corresponding super-peer (in its vicinity). 
\subsection{Protocol Specification}
PJ-Sec facilitates to compute an overlay  ID for a new node (N) in collaboration of three entities (new node ($N$), its friend node ($B$) and vicinity head ($V$)).  The Operation of the proposed scheme consists of three phases, namely, initialization phase, joining phase, and the endorsement phase. as depicted in Figure \ref{fig:layoutproschm} and discussed below. 
\begin{enumerate}
\item
\textbf{Initialization Phase:} Let $\mathbb{G}$ be a group of large prime order $\mathbb{P}$ and $G \in \mathbb{G}$ be the primitive element of $\mathbb{G}$. The system-wide parameters $\langle \mathbb{G}, \mathbb{P} , G\rangle$ 
are available to all entities. The bootstrap node ($B$) and vicinity head ($V$) already exist in the overlay and thus, they have chosen $s_b$ and $s_v$ $ (\in {\leftarrow}{\mathbb{Z}_\mathbb{P}}^*$) as their private key respectively. $PU_{B}$ (= $s_b \boldsymbol{\cdot} G$) and $PU_{V}$ (= $s_v\boldsymbol{\cdot} G$) are the corresponding public keys of B and V.

\item
\textbf{Joining Phase:}
It is assumed that the new node (N) willing to join into the overlay, has all the public parameters and the ID of at least one friend node. The public parameters can be obtained while installing the overlay code, while a friend nodeID can be obtained through a different channel. 

$N$ selects a random number $s_n$ $\overset{\,U}{\leftarrow}{\mathbb{Z}_\mathbb{P}}^*$ as its private key, calculates $PU_{N1}$ ($ = s_n \boldsymbol{\cdot} G$) and 
executes the following steps.
\begin{itemize}
  
\item[Step 1:] $N$ contacts to its friend node ($B$), with a joining request containing its unique ID (such as IP address/ real identity) and its public key as $ \langle IP_N|| PU_{N} \rangle$. 

\item[Step 2:] After receiving the joining request from $N$, $B$ extracts $IP_N$ and $PU_{N}$ and calculates $\beta_1 = (s_b \boldsymbol{\cdot} PU_{N})$. Afterwards $B$ sends  $\langle IP_N||PU_{N1}||\beta_1\rangle$ to $V$. 

\item [Step 3:] Once, V receives the forwarded joining request from $B$, it computes $\gamma_1= (v \boldsymbol{\cdot} \beta_1)$  and $\gamma_2= (v \boldsymbol{\cdot} PU_{B})$.  If forwarded node is not a trusted node, the forwarded ID to generate a nodeID would be rejected. Otherwise, $V$ computes overlay ID for node $N$ ($ID_N$) as:
\begin{equation}
\label{nodeID}
\mathbb{H}( IP_N||PU_{N}||\gamma_1)
\end{equation}
\\ where, $\mathbb{H}$ is a standard hash function like SHA1. 

Pilot V  generates a token ($T_N$)  as shown in Eqn. \ref{token} for $ID_N$, which includes unique real ID of node N ($IP_N$), its public key ($PU_{N}$), friend node overlay ID ($ID_B$), the Pilot ID ($ID_V$) and issuing token time stamp ($T$). V returns the signed token ($T_N$) to $B$ (friend node of $A$) which is signed by the pilot's private key ($s_v$).
\begin{equation}
T_N = Sign_{s_v} \{IP_N, PU_{N}, ID_N, ID_B, ID_V, T\} \label{token}
\end{equation}

\item[Step 4:] B verifies the $ID_N$ using its private key as follows.  It computes $\beta_2 = (s_b \boldsymbol{\cdot} \gamma_2)$ and  $\beta_3 = s_v \boldsymbol{\cdot} s_b \boldsymbol{\cdot} G$. Subsequently, it computes $ID'_N$ as $\mathbb{H}( IP_N||PU_{N}||\beta_2)$. Then, it verifies $ID'_N \? ID_N$. On successful verification B sends $(ID_N || T_N  || \beta_3)$ to N. 

\end{itemize}
\textbf{Endorsement Phase:} 
After receiving the response from $B$, $N$ computes $\alpha_2= (s_n \boldsymbol{\cdot} \beta_3)$ and $ID_N' = \mathbb{H}(IP_N || PU_{N} || \alpha_2)$. Finally, it checks $ID_N' \? ID_N$ to confirm its token ($T_N$) and use it for further communication from other members. 

\end{enumerate}

Later on, while communicating a peer can verify the $ID_N$ from its token. If the verifying peer trusts the vicinity node, it trusts $ID_N$ otherwise computes  the trust through its associated pilot\ vicinity-head.

\begin{figure*}[htb]
\centering
\scalebox{0.9}{
\begin{tabular}{|c c c|}
\Cline{0.5pt}{1-3} 
{\fbox{{\textbf{New Node (N)}}}} & ~~~~~~~~{\fbox{{\textbf{Bootstrap Node (B)}}}} & ~~~~~\fbox{{\textbf{Vicinity Head (V)}}}\\
&&\\ 
 & Private Key $s_b \in {\mathbb{Z}_\mathbb{P}}^*$  & Private Key $s_v \in {\mathbb{Z}_\mathbb{P}}^*$  \\
 & Public Key $PU_{B} = s_b \boldsymbol{\cdot} G$ & Public Key $PU_{V} = s_v \boldsymbol{\cdot} G$\\
&&\\
& {~~~~~~~\fbox{{\it{Joining Phase}}}}&\\
&&\\
Private Key $s_n \in {\mathbb{Z}_\mathbb{P}}^*$ &   &   \\
Public Key $PU_{N} = s_n \boldsymbol{\cdot} G$ &  &  \\
&&\\
\multicolumn{2}{|c}{~~~~~~~$\xrightarrow{\makebox[3.5cm]{$<IP_N || PU_{N}>$}}$} & ~ \\
& $\beta_1= s_b \boldsymbol{\cdot} PU_{N}$ &  \\
& \multicolumn{2}{c|}{~~~~~~~~$\xrightarrow{\makebox[3.5cm]{$<IP_N || PU_{N}|| \beta_1>$}}$}\\

&  & $\gamma_1 = s_v \boldsymbol{\cdot} \beta_1$ \\
&  & $\gamma_2 = s_v \boldsymbol{\cdot} PU_{N}$ \\
&  & $ID_N = \mathbb{H}(IP_N || PU_{N} || \gamma_1 )$ \\
&  & $\tau = (IP_N, PU_{N}, ID_N, ID_B,$ \\
&  & $ ID_V, \gamma_1 , Time)$ \\
&  & $T_N$ = $Sign(s_v, \tau) $ \\

&&\\
& \multicolumn{2}{c|}{~~~~~~~~$\xleftarrow{\makebox[3.5cm]{$(ID_N || \gamma_2   || T_N)$}}$}\\

&&\\
& {~~~~~~~\fbox{{\it{Endorsement Phase}}}}&\\
&&\\
& Verify$(PU_{V}, (Sign(s_v, \tau)))$  &  \\

& Computes $\beta_2=s_b \boldsymbol{\cdot}   \gamma_2 $  and $\beta_3=s_b \boldsymbol{\cdot} PU_{V}$ &  \\
& Computes $ID_N' = \mathbb{H}(IP_N || PU_{N} || \beta_2)$ &  \\
& $ID_N' \? ID_N$ &  \\
&&\\
\multicolumn{2}{|c}{~~~~~~~$\xleftarrow{\makebox[3.5cm]{$<ID_N || \beta_3  || T_N > $ }}$} & ~ \\
&&\\
$\alpha_1= s_n \boldsymbol{\cdot} PU_{V}$ & &  \\
$ID_N' = \mathbb{H}(IP_N || PU_{N} || \alpha_1)$ &  & \\
$ID_N' \? ID_N$ &  & \\
&&\\
\Cline{0.5pt}{1-3}
\end{tabular}}
\scalebox{0.9}{
\begin{tabular}{c}
\multicolumn{1}{p{15cm}}{$IP_N$: Mobile Number or any unique identification of  a new node (N) and $IP_N \in G$. }\tabularnewline
\end{tabular}}
\scalebox{0.9}{
\begin{tabular}{c}
\multicolumn{1}{p{15cm}}{$\mathbb{H}$ = ${\mathbb{Z}_\mathbb{P}}^* \times {\mathbb{Z}_\mathbb{P}}^* \times {\mathbb{Z}_\mathbb{P}}^* \rightarrow {\{0, 1\}}^*$.}\tabularnewline
\end{tabular}}
\caption{Schematic diagram of the proposed scheme.}
\label{fig:layoutproschm}
\end{figure*}
\section{Analysis}
\label{analysis}
\subsection{Design Requirement Analysis}
\label{desgReq}
Our proposed nodeID assignment PJ-Sec achieves the following features.
\begin{itemize}
\item {Uniqueness:} 
In our scheme, each user uses its own unique identity ($IP_N$) as its mobile number. It assumes that smart-phones are the component of the cellular P2P overlay; therefore, they are verified by its unique identity to stop generating many overlay IDs for a single mobile. Each unique overlay ID is wrapped with its unique identification to generate a unique nodeID and its corresponding token. This token can be verified at each communicating peer to prevent the Sybil IDs. So each ID is assumed to be unique.
\item {Stability:} 
The proposed scheme collaborates an existing member peer and the head peer to offer stability among overlay IDs.  Therefore, a single node fails to change the nodeID.  
\item {Joint Management:}   
In our mechanism, each new node concatenates its public key with own unique identity ($IP_N$) and sends to a friend node ($B$) with joining request in the overlay. Further, this information forwards to vicinity head ($V$) after concatenating node $B$’s public key. A token ($T_N$) is generated and verified by the friend node before the final assignment of new ID. Moreover, the nodeID (Eqn. \ref{nodeID}) is computed as the collaborative effort of three different nodes $N$, $B$, and $V$.
\item {Verifiable:} 
In our approach, the generated token (Eqn. \ref{token}) can verify to the  nodeID $ID_N$ and further, It can be verified by any member node.  
\item {Traceability:}  
Our scheme traces any node with the help of six parameters of the token $T_N$, i.e., 
\begin{gather*}
 (IP_N, \alpha_1, ID_N, ID_B, ID_V, time)
\end{gather*}

As described earlier in the protocol specification section. In the verification phase, the ID ($ID_N$) can be checked with its real identity ($IP_N$) along with the collaborated nodeIDs and the validity concerning issuing time.  
\item {Revocability:} 
The proposed scheme achieves this requirement at a pilot with the help of received token $T_N$. Since the pilot keeps the information of its member MSs, the forge IDs are detected and the corresponding token would be revoked.  
\item {Uniformity:}  
In our proposed mechanism, the joining request of the new node would be forwarded to another vicinity head which is connected to WiFi in the case of more than ($2^p$) smart-phones within p-bit virtual space of a vicinity head.
\item {Scalability:}
Large mobile P2P overlay is virtually divided into vicinity range to make less computation and inexpensive communication (D2D/Bluetooth/WiFi) during ID generation in our proposed mechanism. So, the overlay structure supports huge new peers to join with respective vicinity without using Internet cost and much energy consumption. 
\end{itemize}
\subsection{Security Analysis}
This sub-section analyses the security aspect of PJ-Sec from different attacks related to the nodeID assignment. 
\begin{itemize}
    \item {Sybil Attack:}
 In PJ-Sec, each ID is a  function of its $IP_N$, $PU_N$ and $\gamma_1$, which is signed by the vicinity head and endorsed by the friend node in the token. Therefore, a malicious node fails to generate Sybil IDs without compromising a friend node and the Vicinity head.

\item{Eclipse Attack:}
Any node cannot generate own preferable IDs to place itself near to a target node due to the collaborative nodeID generation process and a random hash digest. Therefore, the attackers are not able to select any victim peer to launch an Eclipse attack.

\item{MITM Attack:}
It is not feasible for an attacker to successfully execute the  node assignment phase of PJ-Sec due to the verification operation at each stage. 
\end{itemize}

\subsection{Formal security verification}
\label{avispa}
Our proposed secure nodeID assignment protocol is verified/validated using a state-of-the-art tool (AVISPA~\cite{armando2005avispa}), which provides automatic security verification and analyzes the specified security goals to measure whether the mechanism is SAFE or not.  To examine a protocol through AVISPA, it is to be codded in High-Level Protocol Specification Language (HLPSL) and integrated through back-end servers. 
These servers are responsible for providing automatic security verification and analysis for the HLPSL specification codes after getting the intermediate form (IF) of the code using  $hlpsl2if$ translator. 

The present version of AVISPA supports four back-end servers which can be integrated with HLPSL. The first back-end is  ``On-the-Fly Model-Checker" (OFMC), which explores the network state through the demand-driven way. Second, CL-AtSe back-end translates the transition relation of a protocol specification in the intermediate format (IF) into a set of constraints to find whether attacks have been imposed or not. SAT is the third back-end based on the model checker, which generates and feds propositional formulae into SAT solver. The found module in the SAT process would be translated back into an attack to analyze the specified security goals. TA4SP is the fourth back-end which stands for ``Tree Automata based on Automatic Approximation for the Analysis of Security Protocols". It approximates the intruder's knowledge using regular tree language.  

In AVISPA, each participant assigns a role to play with some initial parameters and communicates with others through the channel (Dolev-Yao (dy) model ~\cite{dolev1983security}) during the protocol execution. In this, the declaration channel (i.e., dy)  may be secure or insecure. The intruders assume to play a legitimate role during the simulation run.  After the successful execution, OUTPUT FORMAT (OF) is generated to describe the security analysis under the given conditions.       

\begin{figure*}[!th]
    \centering
    \includegraphics[width=5.5in]{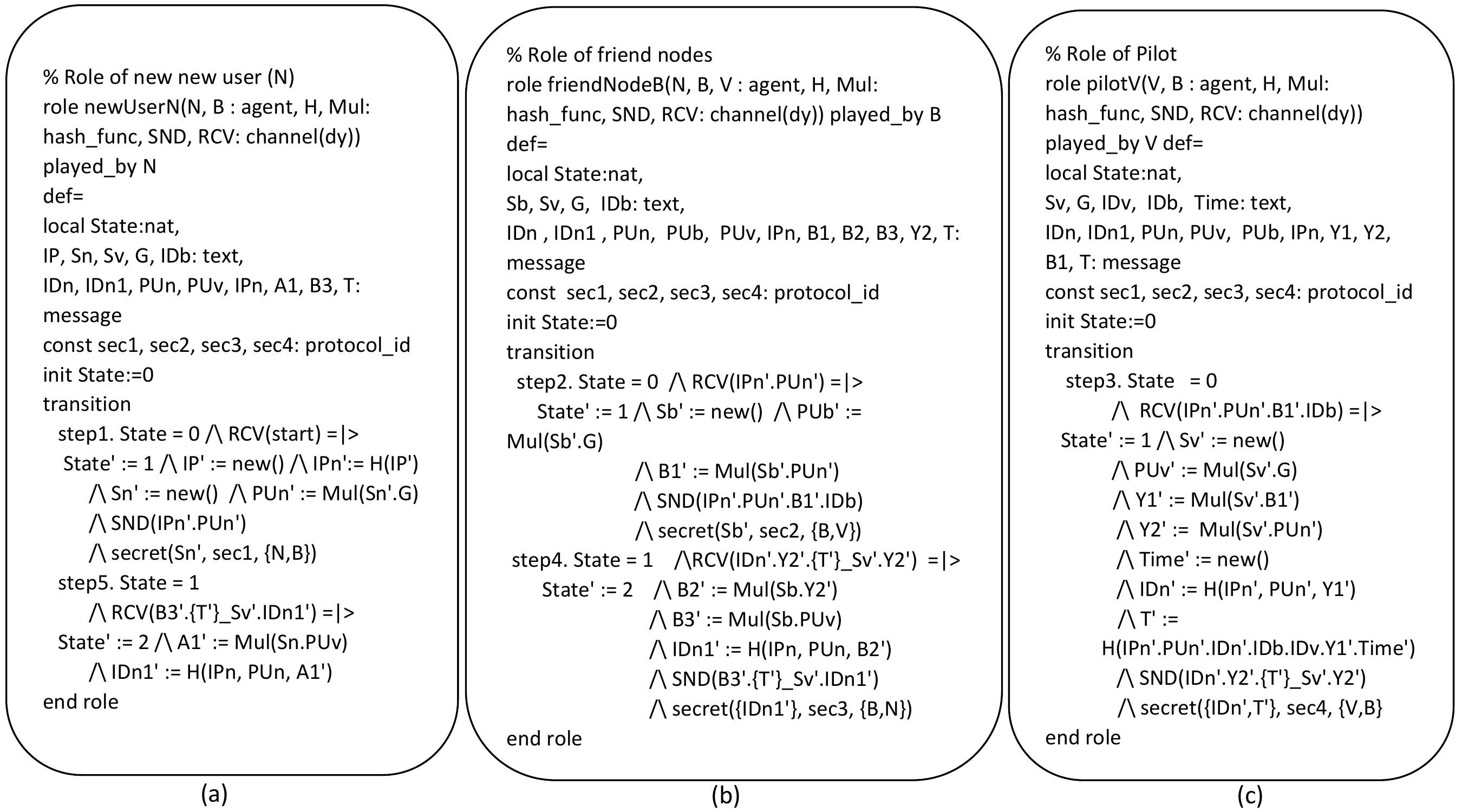}
    \caption{Role specification: a) for new node N,  b) for friend node B and c) for pilot V.}
    \label{fig:Roles}
\end{figure*}

\begin{figure}[!th]
    \centering
    \includegraphics[width=2.3in]{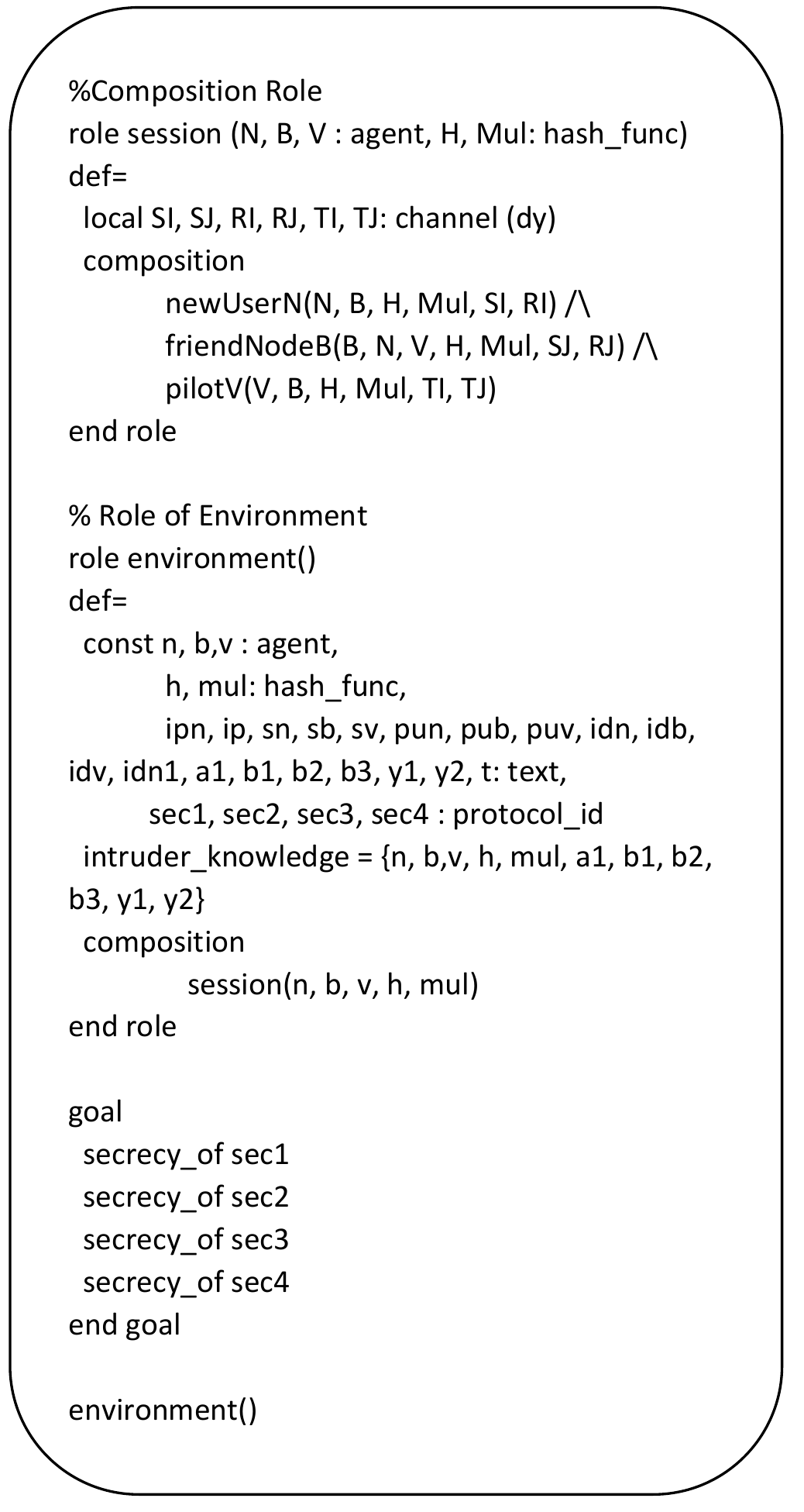}
    \caption{The session, goal and environment specification.}
    \label{fig:GoalEnvironment}
\end{figure}

\begin{figure}[!th]
    \centering
    \includegraphics[width=3.4in]{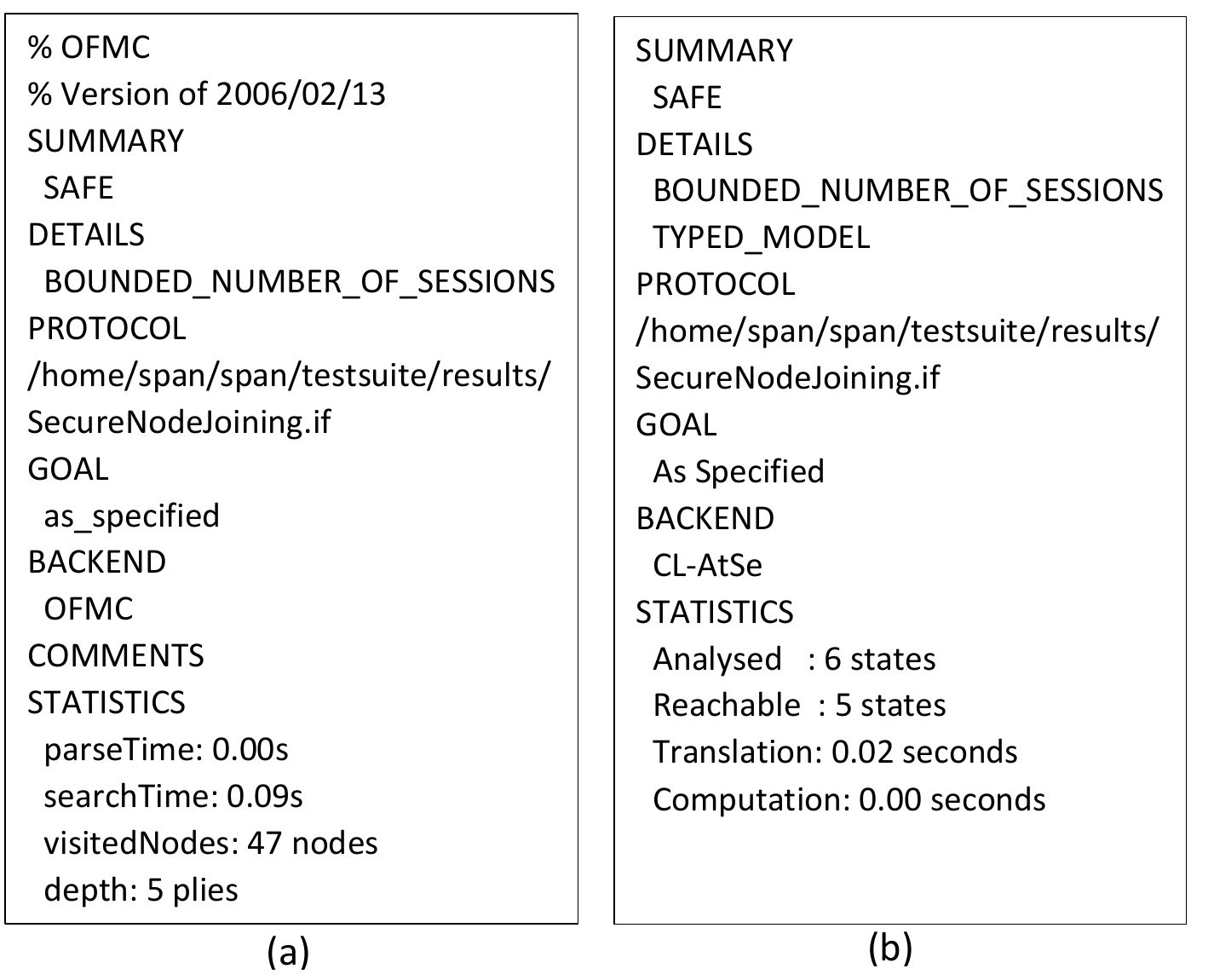}
    \caption{Simulation results by a) the OFMC back-end and b) the SL-AtSe Back-end.}
    \label{fig:results}
\end{figure}

All roles specification of the players (i.e., new node $N$, friend/bootstrap node $B$, and pilot $V$) in the proposed protocol using HLPSL language are depicted in Figures \ref{fig:Roles} (a), (b), and (c) respectively. The operators $secret()$ is used to verify the secrecy of communication between players.
In this verification, the intruders are provided with the knowledge of the players with their overlay Ids and common system parameters. Figure ~\ref{fig:GoalEnvironment} defines the session, security goals, and the environment for our proposed protocol.  
The results are shown in Figures ~\ref{fig:results} (a) and (b) are obtained after execution of the security test at OFMC and CL-AtSe back-end respectively. Both the results confirm that our scheme satisfies authentication goals for participants. 
%
\section{Implementation and Result Analysis}
\label{dis}
We implemented the schemes using pairing-based cryptography (PBC) library in an Intel Core i5 2.30GHz processor laptop PC with 64 bits OS type Ubuntu 16.04 LTS system to calculate the computational cost, whereas MATLAB R2016a is used to determine the consumed network bandwidth during node generation process.
\subsection{Computational Cost}
%

Figure ~\ref{fig:netTraf} compares the computational overhead of our proposed PJ-Sec and the existing secure node ID assignment through key-based mechanisms. Our scheme took 0.0357735 seconds while RIAPPA~\cite{caubet2014riappa} and IAP ~\cite{caubet2013securing} (existing protocols) took 0.124545 seconds and 0.041516 seconds respectively.
\begin{figure}[!t]
    \centering
    \includegraphics[width=3.2in]{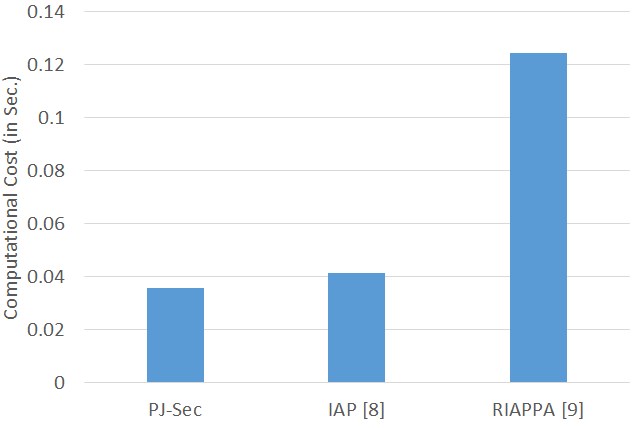}
    \caption{Computational cost due to new node ID assignment.}
    \label{fig:netTraf}
\end{figure}
We compared the proposed scheme also with the existing schemes on the basis of intensive operations involved and summarized in Table ~\ref{comput}. The comparison shows (Figure~\ref{fig:netTraf}) that our scheme has significantly less operations than others. 
\begin{table}
\caption{Summary of cryptographic operations.}
\label{comput}  
\centering     
\begin{tabular}{|l|c|c|c|c|}

\hline  Security Protocols & Mul & Add & Hash & Inverse  \\
\hline   Our Proposed & \cellcolor{green!25} 23 & 12 &  3 & \cellcolor{green!25} 1 \\
\newline Protocol (PJ-Sec) &  \cellcolor{green!25} &  &  & \cellcolor{green!25} \\

\hline  IAP ~\cite{caubet2013securing}  & 52 & 10 & 3 & 8 \\

\hline  RIAPPA ~\cite{caubet2014riappa}  & 147 & 25 & 3 & 21 \\
\hline
\end{tabular} 
\end{table}
\subsection{Bandwidth Consumption}
To know the bandwidth consumption during secure ID assignment, we use the performance estimation of network Traffic model ~\cite{tetarave2018v} to compare our proposed protocol with the existing mechanisms. In this, the available bandwidth ($R_0$) is set randomly in the range [0.75$R_0$, 1.25$R_0$]. MSs and pilots are set at the maximum bandwidth 0.1$R_0$ ($C_p$), while it is 1.0$R_0$ ($C_w$) for the eNodeB and backbone network. We set $R_0$ = 100 kbps to approximate it to a realistic speed during a file upload. 
During secure node joining process, the protocols use Shannon capacity formula (S(bits/s/Hz)) as in~\cite{mogensen2007lte}. It captures the time-varying capacity of the wireless channel. 
\begin{equation}
S(bits/s/Hz)= BW_{eff}.\eta .\log_2( 1+ SNR / SNR_{eff} ) 
\label{ShannonCap}
\end{equation}
where $BW_{eff}$ is the efficiency of LTE  to approximate for the system bandwidth ($R_0$).  $SNR_{eff}$ approximates for the SNR implementation of the efficiency of LTE. The factor $\eta$ is a correction factor which is considered as one. 
In our simulation,  we used the values $\eta$ = 0.9 and $SNR_{eff}$  = 1.23 which is the best fit to the link adaptation curve~\cite{mogensen2007lte}.  It is also considered in the network that channel fitting takes an upper limit of $S$ according to the hard spectral efficiency given by modulation and coding set.  For the single stream case, coding set is considered as 4/5, e.g., 64QAM.  

For sending joining request in centralized solutions such as RIAPPA~\cite{caubet2014riappa} protocol, new nodes have to communicate internal TTP and then, the TTP communicates with external TTP for generating the ID. So, total communication ($TC_{RIAPPA}$) for generating a new key within RIAPPA would be $3*C_w$. On the other hand, IAP~\cite{caubet2013securing} has to communicate with single TTP to assign a new ID which takes $2*C_w$ communications. For distributed solutions as in  Dinger \textit{et al.} ~\cite{dinger2006defending}, a new node ID assignment mechanism performs the trust among $(r/2)$ number of nodes and they can be transferred the request through eNodeB or WiFi. So, total communication ($TC_{DistributedSol}$) for assigning a secure ID for a new member within this distributed solution would be $1.5(r*C_p)$. In our proposed mechanism, new ID generates securely with the collaborative effort of vicinity head and an existing node within its vicinity. Therefore, total communication ($TC_{PJ-Sec}$) for generating a secure nodeID would be $4*C_p$. 
In the simulation, We analyze the capacity required to serve these requests to assign new IDs up to 500 new nodes. The result shows that the total network trafficking is significantly lower in our proposed security scheme (PJ-Sec) than other security protocols as depicted in Figure \ref{fig:netTraf}.
\begin{figure}[!t]
    \centering
    \includegraphics[width=3.2in]{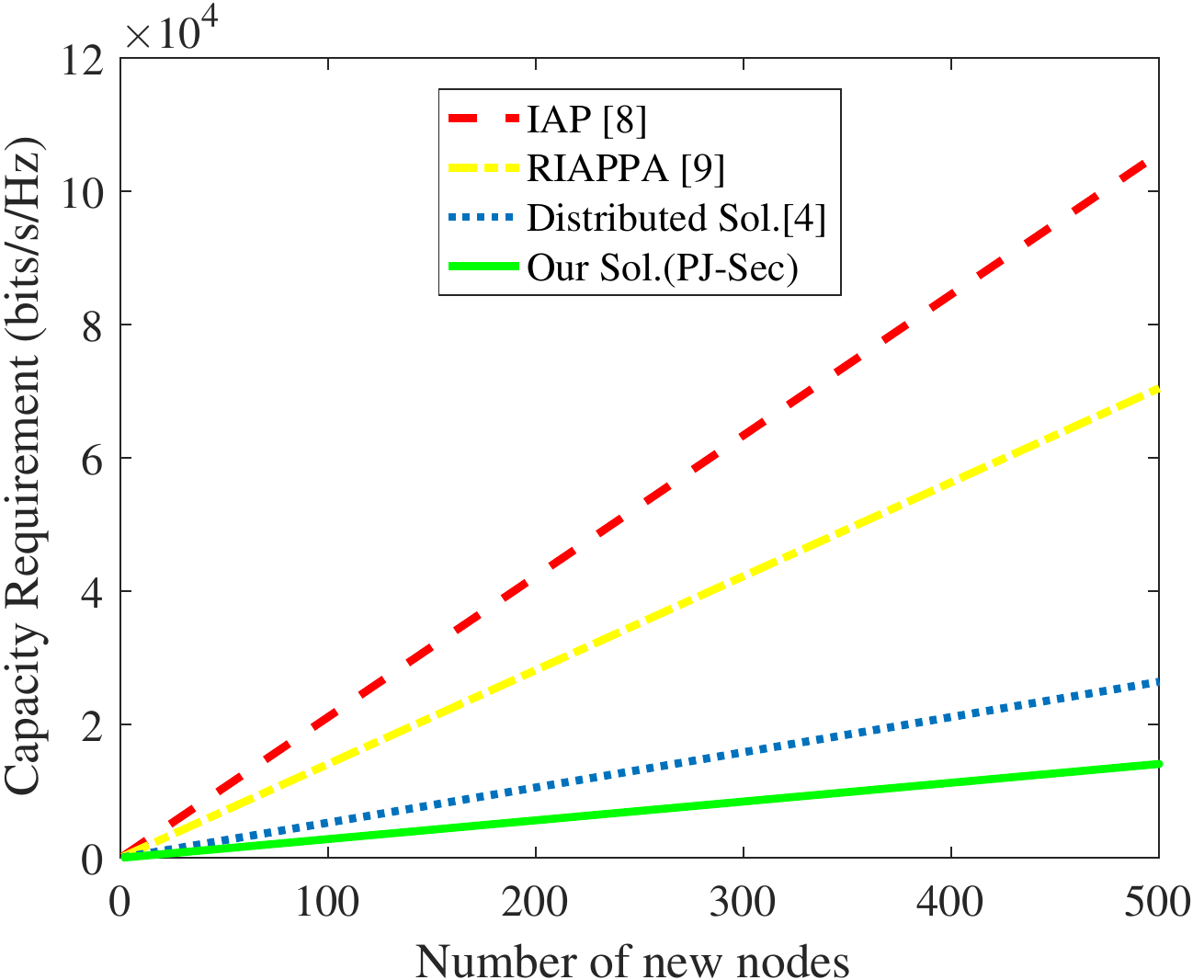}
    \caption{Bandwidth consumption due to new node ID assignment.}
    \label{fig:netTraf}
\end{figure}

\begin{figure}[!t]
    \centering
    \includegraphics[width=3.2in]{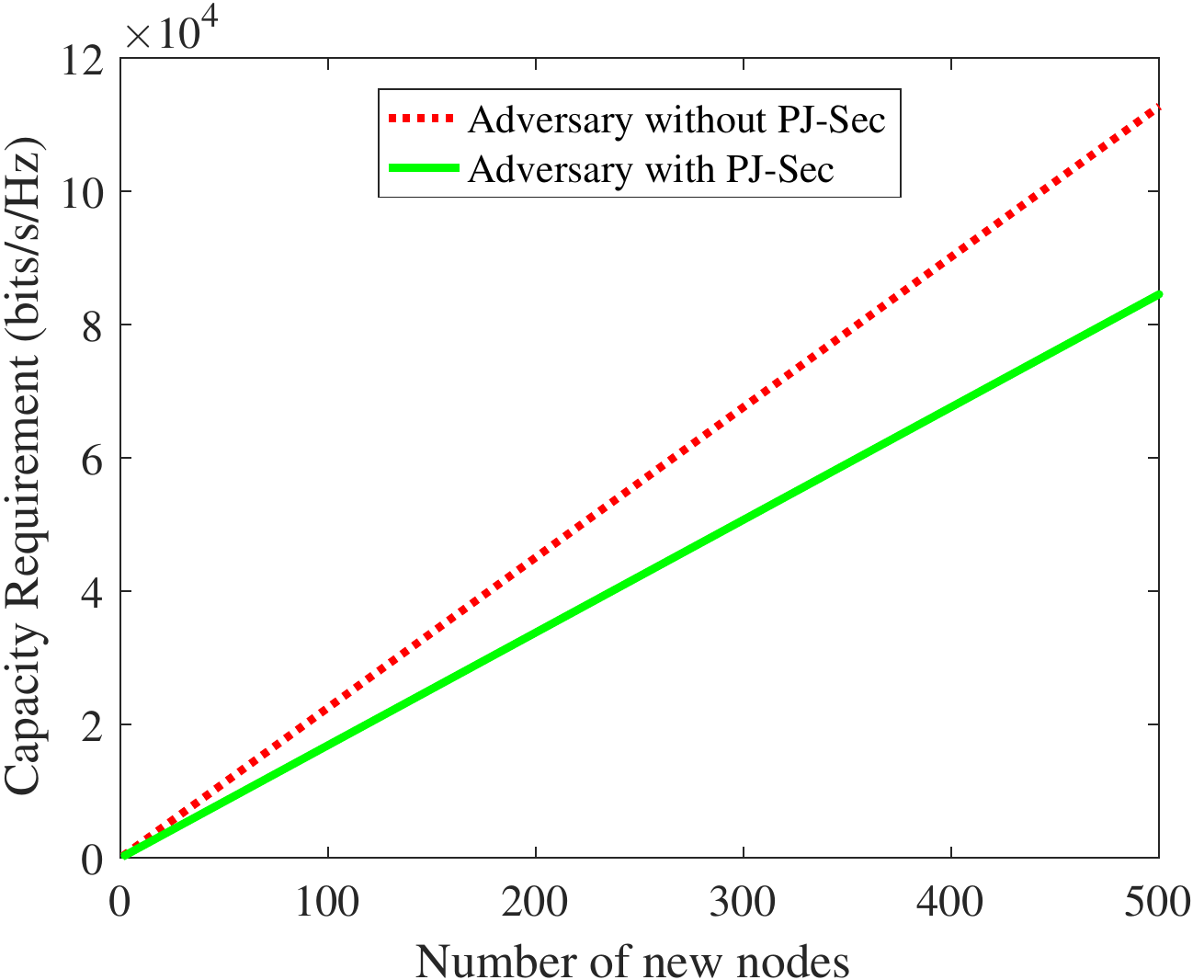}
    \caption{Bandwidth consumption with and without implementing the proposed protocol PJ-Sec, where pilot-peer group consists of 64 nodes and forge Id domain space is 3-bit (fixed).}
    \label{fig:infNoninf}
\end{figure}

\begin{figure}[!t]
    \centering
    \includegraphics[width=3.2in]{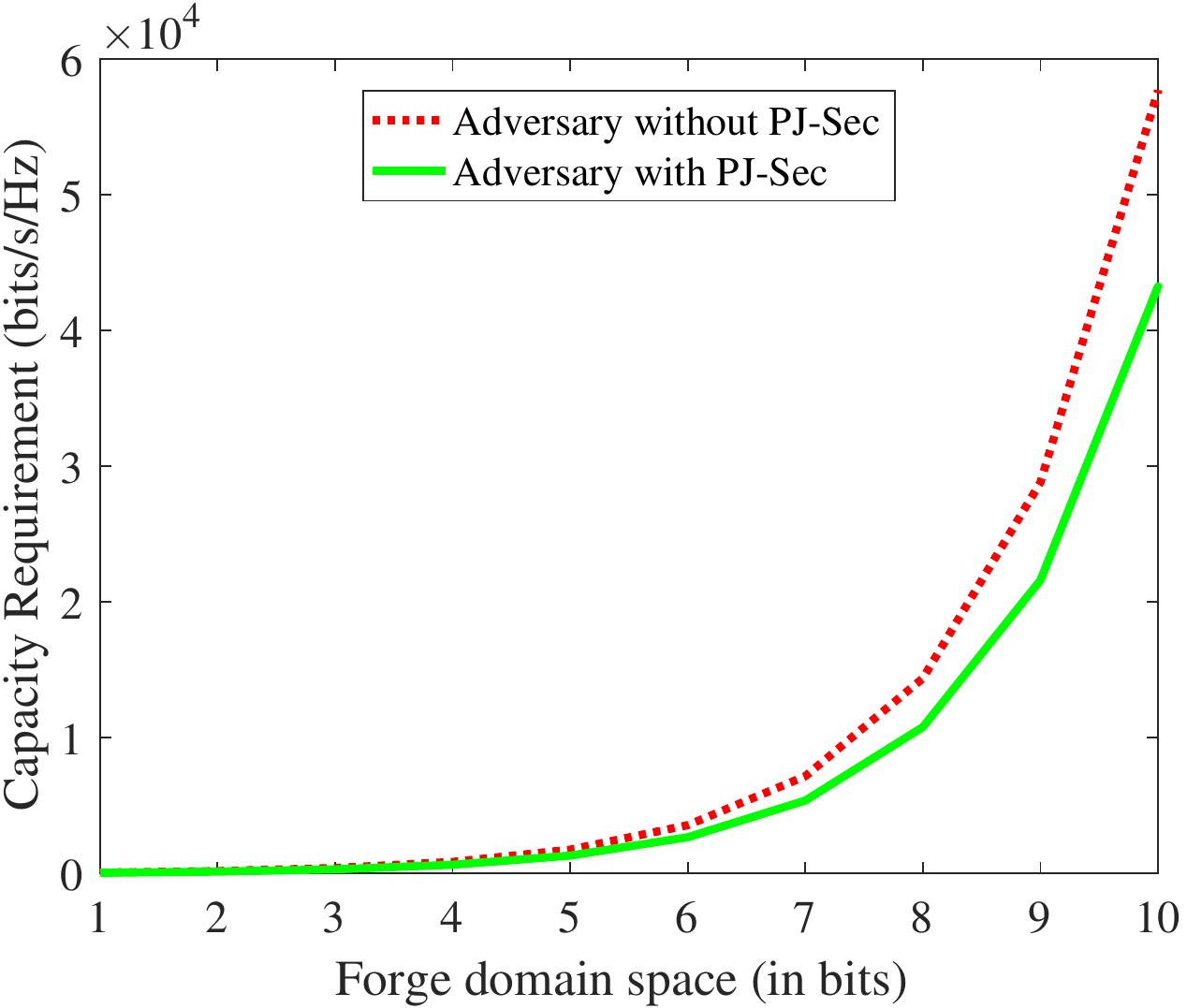}
    \caption{Effect of Sybil Id domain space from 1 to 10-bit, where adversary mitigates with the proposed protocol PJ-Sec or having no mitigation mechanism.}
    \label{fig:forgeDomain}
\end{figure}

Further, we analyze the bandwidth consumption effect in the presence of adversaries with or without mitigation through our proposed solution (PJ-Sec). Figure ~\ref{fig:infNoninf} compares the bandwidth consumption to generate Sybil (or forge) Ids within a fixed 3-bit domain and the reduced consumption using our security mechanism PJ-Sec. It happens due to each unsuccessful verification (in PJ-Sec), at mediator (or friend) nodes, stops new Id generation process. Figure \ref{fig:forgeDomain} shows the effect of an expansion of the forge Id domain space from 1 to 10-bit. The proposed protocol requires significantly lesser bandwidth as the forge Id domain space is expended.
\section{Conclusion}
\label{con}
In this work, we propose a secure node joining protocol, called PJ-Sec tailoring to DHT-based mobile P2P. The proposed PJ-Sec protocol can be utilized to tune other existing P2P mechanisms as well. To defend against Sybil attack, PJ-Sec provides nodeID as a collaborative effort of existing overlay peer (friend node) and vicinity head.  PJ-Sec is formally proved using the well-known AVISPA verification tool and shown to be secure. The performance of the scheme in regards to computation and bandwidth cost is analyzed and observed to be highly efficient when compared with the existing schemes.  Implementing the whole technique on a smart-phone is a part of our ongoing work.

\section*{Acknowledgement}
This work is funded by the E-security Division, Ministry of Electronics
and Information Technology, Government of India, through the project grant
number 12(7)/2015-ESD.

\bibliographystyle{unsrt}  
\bibliography{reference} 

\end{document}